\documentclass[vci-paper]{vciart}
\usepackage{amssymb}
\usepackage{graphicx}

\begin{document}
\begin{frontmatter}

\title{Reading a GEM with a VLSI pixel ASIC used as a direct charge collecting anode}

\author[A]{R. Bellazzini},
\author[A]{F. Angelini},
\author[A]{L. Baldini},
\author[A]{F. Bitti},
\author[A]{A. Brez},
\author[A]{L. Latronico},
\author[A]{M.M. Massai},
\author[A]{M. Minuti},
\author[A]{N. Omodei},
\author[A]{M. Razzano},
\author[A]{C. Sgr\`{o}},
\author[A]{G. Spandre},
\author[B]{E. Costa},
\author[B]{P. Soffitta}

\address[A]{INFN Pisa, Via Buonarroti 2, 56127 Pisa, Italy}
\address[B]{Istituto di Astrofisica Spaziale del CNR, Area di Ricerca di Roma, V. Fosso del Cavaliere, 00131 Rome, Italy}

\begin{abstract}
In MicroPattern Gas Detectors (MPGD) when the pixel size is below 100 $\mu$m and the number of pixels is large (above 1000) it is 
virtually impossible to use the conventional PCB read-out approach to bring the signal charge from the 
individual pixel to the external electronics chain. For this reason a custom CMOS array of 2101 active pixels with 
80 $\mu$m pitch, directly used as the charge collecting anode of a GEM amplifying structure, has been developed 
and built.
Each charge collecting pad, hexagonally shaped, realized using the top metal layer of a deep submicron VLSI technology is 
individually connected to a full electronics chain (pre-amplifier, shaping-amplifier, sample \& hold, multiplexer) which 
is built immediately below it by using the remaining five active layers. 
The GEM and the drift electrode window are assembled directly over the chip so
the ASIC itself becomes the pixelized anode of a MicroPattern Gas Detector. 
With this approach, for the first time, gas detectors have reached the level of integration and resolution typical of solid state 
pixel detectors. Results from the first tests of this new read-out concept are presented. 
An Astronomical  X-Ray Polarimetry application is also discussed.
\end{abstract}
\end{frontmatter}

\section{Introduction}
The most interesting feature of the Gas Electron Multiplier (GEM) is the possibility of full decoupling of the charge amplification
 structure from the read-out structure. In this way both can be independently optimized. Indeed, by organizing the read-out plane in a 
multi-pixel pattern it is possible to get a true 2D imaging capability. 
At the same time a high granularity of the read-out plane would also allow to preserve the intrinsic resolving power of the 
device and its high rate capability that otherwise would be unavoidably lost by using a conventional projective read-out approach.  
 However, when the pixel size is small (below 100 $\mu$m) and the number of pixels is large (above 1000) it is virtually impossible 
to bring the signal charge from the individual pixel to a chain of external read-out electronics even by using an advanced, fine-line,
 multi-layer, PCB technology. The fan-out which connects the segmented anodes collecting the charge to the front-end electronics 
is the real bottleneck. Technological constraints limit the maximum number of independent electronics channels that can be brought 
to the peripheral electronics. Furthermore,  the crosstalk between adjacent channels and the noise due to the high input capacitance to the 
preamplifiers become not negligible.
 In this case, it is the electronics chain that has to be brought to the individual pixel. 
We have implemented this concept by developing and building a CMOS VLSI array of 2101 pixels with 80 $\mu$m pitch which is used directly 
as the charge collecting anode of the GEM. A description of the read-out ASIC for a MPGD and of its advantages is given in the next section. 
 Section 3 describes the coupling of the chip die to the amplifying electrode, the assembly of the full detector 
and the results of laboratory tests obtained with a 5.9 keV X-ray source. The use of this new detection concept for Astronomical X-Ray 
Polarimetry and other applications are discussed in the last section.

\section{The CMOS VLSI chip}

\begin{figure}[ht]
\begin{center}
\includegraphics[width=10cm]{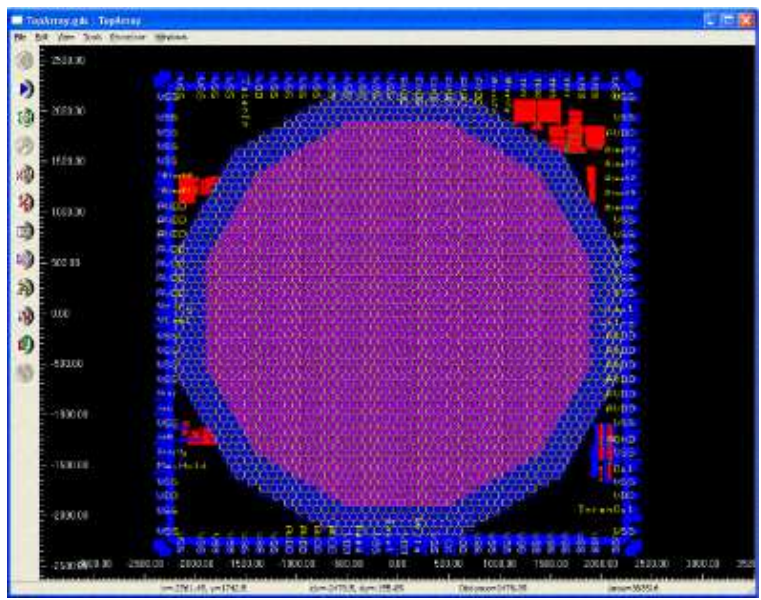}
\caption{The actual ASIC layout as seen from the top (active pixels are pink, guard-ring and I/O pads are blue)}
\label{fig:1}
\end{center}
\end{figure}
\begin{figure}[ht]
\begin{center}
\includegraphics[width=10cm]{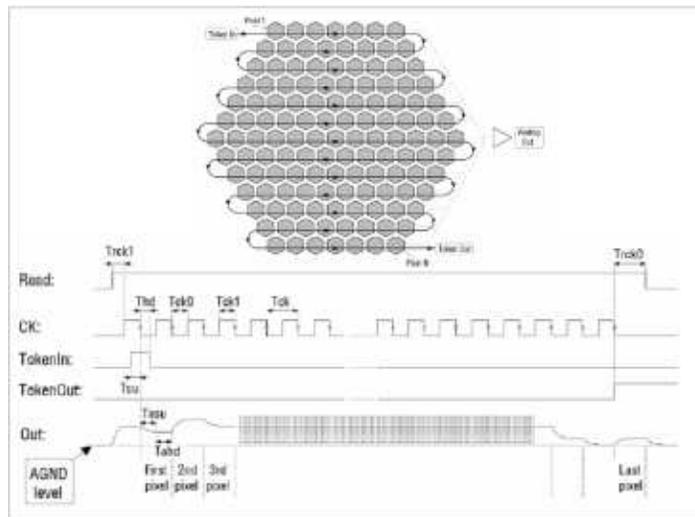}
\caption{Serial read-out architecture}
\label{fig:3}
\end{center}
\end{figure}
\begin{figure}[ht]
\begin{center}
\includegraphics[width=10cm]{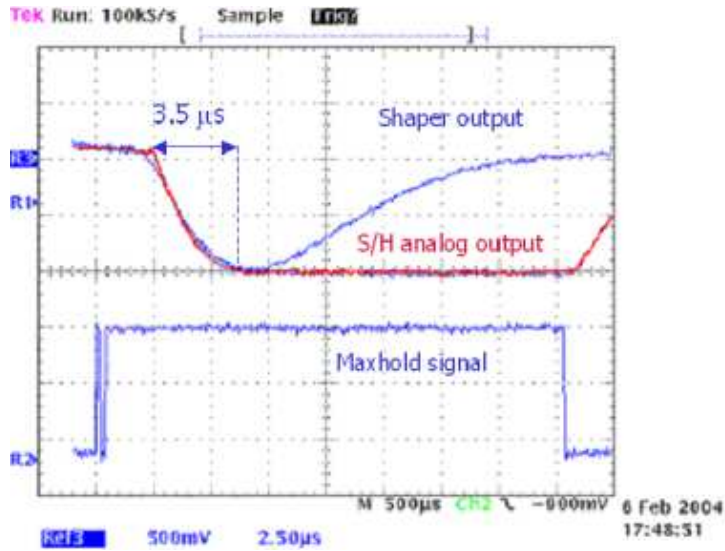}
\caption{Shaped output signal in the two operating modes:{\em Track} mode (MaxHold off) and {\em Peak \& Hold} (MaxHold on)}
\label{fig:4}
\end{center}
\end{figure}
\begin{figure}[ht]
\begin{center}
\includegraphics[width=10cm]{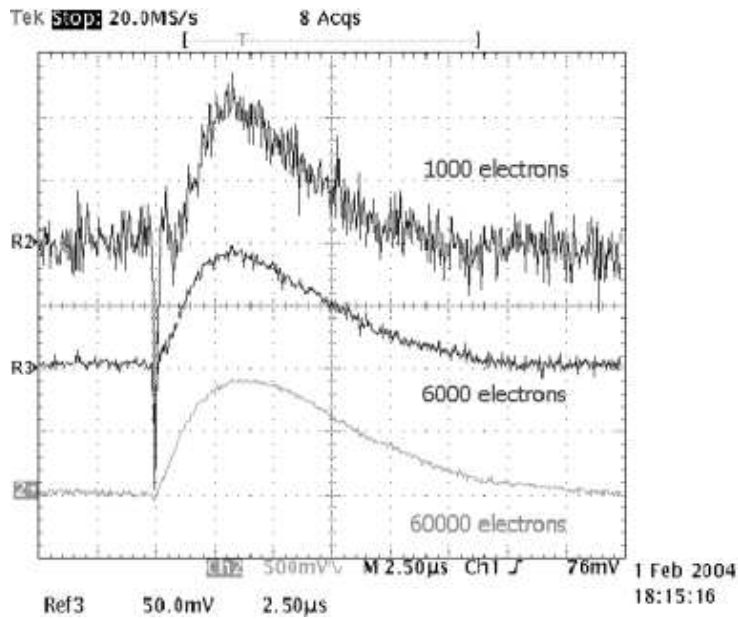}
\caption{A photo of the shaped signals for three different value of input charge}
\label{fig:5}
\end{center}
\end{figure}
\begin{figure}[ht]
\begin{center}
\includegraphics[width=10cm]{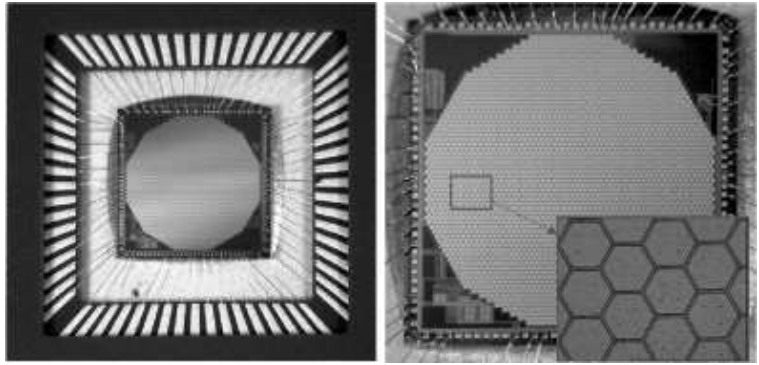}
\caption{A photo of the chip bonded to the CLCC68 ceramic package and a zoomed view of the pixel matrix}
\label{fig:6}
\end{center}
\end{figure}

A drawing of the ASIC layout as seen from the top metal layer is shown in fig.\ref{fig:1}. 
The active matrix in pink  is surrounded by 
a passive guard ring of 3$\div$4 pixels set to the same potential of the active pixels. The chip has been realized using a 0.35 $\mu$m 
3.3 V CMOS technology. 
No specific ESD protection other than the parassitic capacitance of the drain-to-bulk junctions have been foreseen for the pixel pads.
The table below summarizes the electrical characteristics of the chip.
\begin{table}[ht]
\begin{center}
\begin{tabular}{|c|c|}
\hline
\multicolumn{2}{|c|}{\bf ASIC Operating Characteristics}\\
\hline
Parameter                                          & Limit value (unit) \\
\hline
Max supply voltage (Vdd-Vss)                       & 3.6 (V)      \\
Min supply voltage (Vdd-Vss)                       & 3.0 (V)      \\
Max voltage at any pin                             & Vdd+0.3 (V) \\
Min voltage at any pin                             & Vdd-0.3 (V) \\ 
Continuous total dissipation (T $<$60 $^\circ$C)   & 200 (mW)     \\
ESD tolerance of standard I/O                      & 1 (kV) \\
ESD tolerance of pixel pads                        & 100 (pC) \\
Analog output voltage                              & Vss+0.3$\div$ Vdd-0.5 (V) \\
Analog output impedance (0.5V$<$Vout$<$2.5V)        & 5 ($\Omega$) \\
Integral non linearity                             & 0.5 (fC) \\
Average input sensitivity                          & 100 (mV/fC) \\
Pixel-to-pixel gain variation                      & 5 (\%) \\
Recovery time after hold                           & 100 ($\mu$s) \\
Analog calibration input sensitivity, Qin/(Vtest-Vss)  & 10 (fC/V) \\
\hline
\end{tabular}\label{tab} 
\end{center}
\end{table}
Each microscopic pixel is fully covered by a hexagonal metal electrode realized using the top layer of a 6 layers CMOS technology.
Each pad is individually connected to a full chain of {\em nuclear type} electronics (pre-amplifier, 
shaping-amplifier, sample \& hold, multiplexer) which is built immediately below it by making use of the remaining five active layers.  
\begin{figure}[ht]
\begin{center}
\includegraphics[width=10cm]{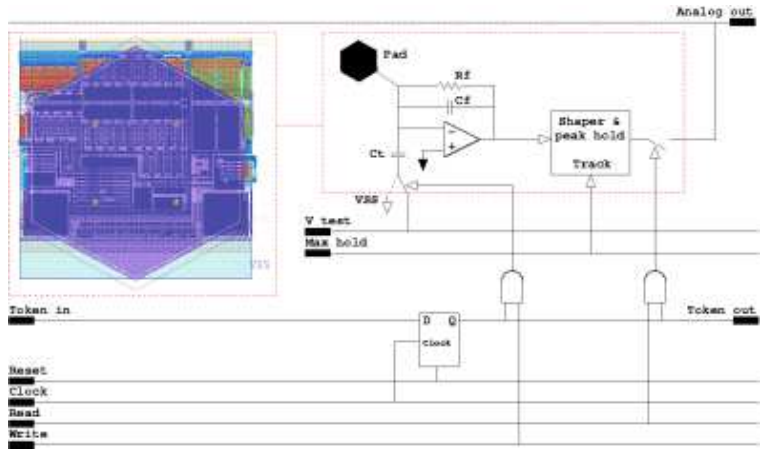}
\caption{A drawing of the pixel layout with underlying electronics and its simplified equivalent electronic scheme}
\label{fig:2}
\end{center}
\end{figure}
Fig.\ref{fig:2} shows the layout and the simplified equivalent scheme for one pixel. 
Upon activation of an external asynchronous trigger (in our case provided by amplifying and discriminating the fast signal obtained 
from the top GEM electrode) and within a 10 $\mu$s window the automatic search of the maximum of the shaped signal starts. 
If the {\em MaxHold} signal is set, the maximum is held for subsequent read-out which is accomplished by sequentially connecting the 
output of each pixel to a common analog bus (fig.\ref{fig:3}). 
A pixel is selected by introducing a token into the shift register and can be electrically stimulated at the rising edge of the 
{\em Write} signal, injecting a charge {\em -Qin} (10fC/V typical response) proportional to the voltage difference betwen {\em Vtest} and {\em Vss}.
Tokens are shifted one cell forward at the falling edge of the input clock.
If several tokens are present in the shift register then the analog output corresponds to the sum of the selected pixels, up to the 
saturation level of $\sim 30fC$. A useful feature of the chip is the possibility to work both in {\em Hold} or  {\em Track} mode.
The shaped pulse from a pixel can be individually observed at the analog out by keeping 
the  {\em MaxHold} signal low.  Fig.\ref{fig:4} shows the signals of a single strobed pixel observed on the digital scope in the 
two different operating modes: {\em Tracking} mode (MaxHold off) and {\em Peak \& Hold} (MaxHold on in red).
Requested specifications for the ASIC prototype are:
\begin{itemize}
\item low noise (typical ENC $\sim$ 100 electrons at 0.1 pF input capacitance),
\item $\sim$ 3.5$\mu$s shaping time,
\item 60$\mu$W typical power consuption per pixel,
\item 5MHz maximum system clock (i.e. serial analog read-out at 200 ns/pixel corresponding to $\sim$ 400$\mu$s total 
read-out time for 2100 pixels),
\item 0.2-20 fC dynamic range.
\end{itemize}
In fig.\ref{fig:5} three different shaped signals obtained injecting a charge of 1000, 6000 and 60000 electrons, respectively, 
in the calibration capacitance are shown. 
This read-out approach has the advantage, respect to similar ones (TFT (\cite{Quote1} or CCD (\cite{Quote2} read-out) of being fully asynchronous 
and externally triggerable. Furthermore it supplies a complete analog information of the collected charge allowing to image
the energy deposition process of the absorbed radiation.
A photo of the actual ASIC bonded to a ceramic CLCC68 package and a zoom over the hexagonal pixel pattern is shown in fig.\ref{fig:6}.

\section{The MPGD assembly}

\begin{figure}[ht]
\begin{center}
\includegraphics[width=10cm]{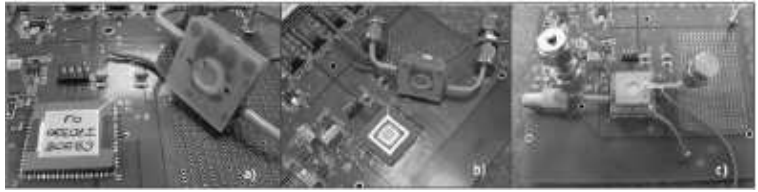}
\caption{Assembly phases of the MPGD over the chip: a) all the mechanical details of the top section of the detector are glued together
while the chip is still protected by a metallic cover, b) the chip is exposed and the mechanics glued upon it, c) the MPGD is ready for test}
\label{fig:7}
\end{center}
\end{figure}

\begin{figure}[h]
\begin{center}
\includegraphics[width=10cm]{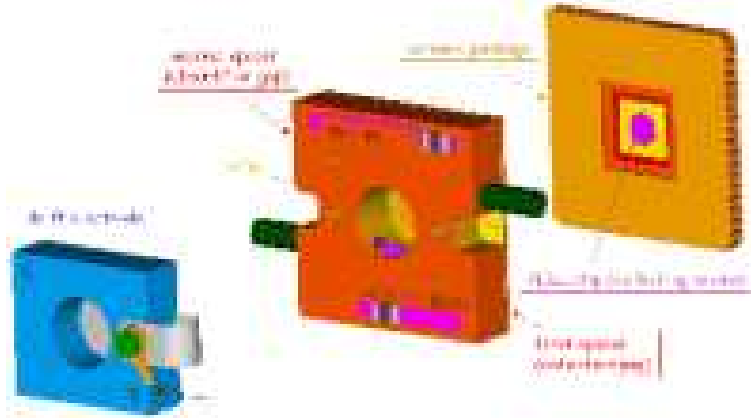}
\caption{Exploded view of the detector assembled over the VLSI ASIC}
\label{fig:8}
\end{center}
\end{figure}

\begin{figure}[h]
\begin{center}
\includegraphics[width=10cm]{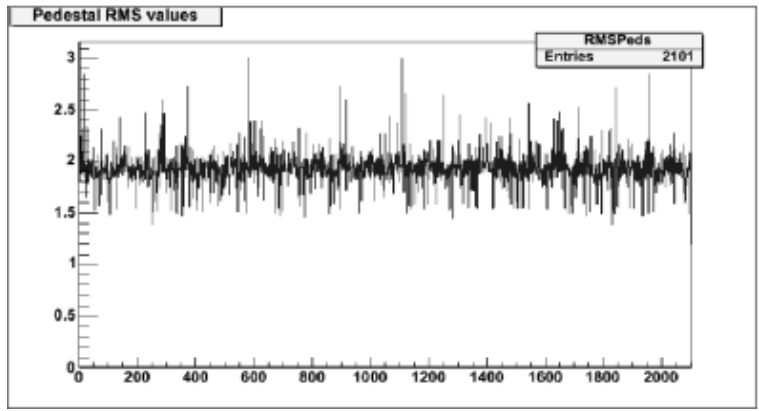}
\caption{Noise measurement: rms value of the pedestals distribution for each electronics channel}
\label{fig:9}
\end{center}
\end{figure}

\begin{figure}[h]
\begin{center}
\includegraphics[width=7cm]{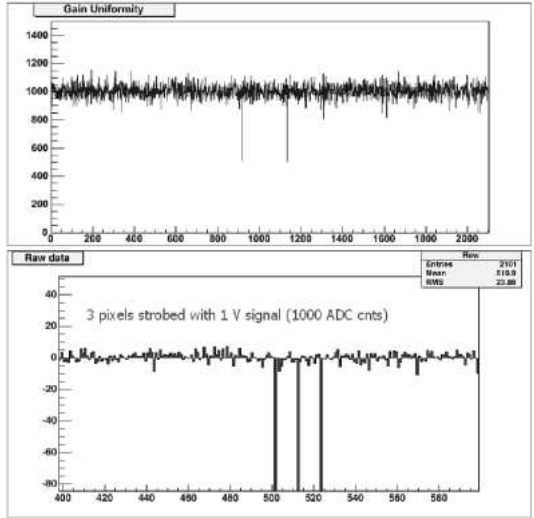}
\caption{Gain uniformity (a) and Xtalk measurement (b)}
\label{fig:10}
\end{center}
\end{figure}

\begin{figure}[h]
\begin{center}
\includegraphics[width=10cm]{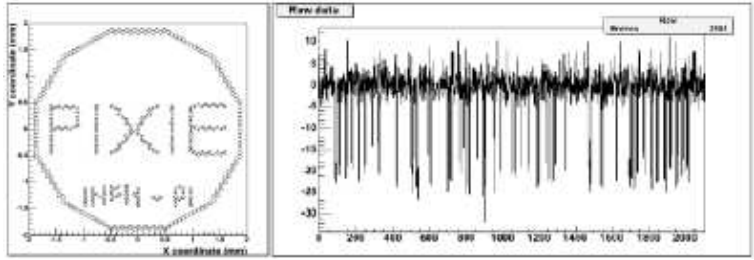}
\caption{Response of a set of selected pixels to 20 mV signal (3 $\sigma$ cut on the noise). 
The corresponding raw signals (pedestal subtracted) are shown on the left}
\label{fig:11}
\end{center}
\end{figure}

\begin{figure}[ht]
\begin{center}
\includegraphics[width=10cm]{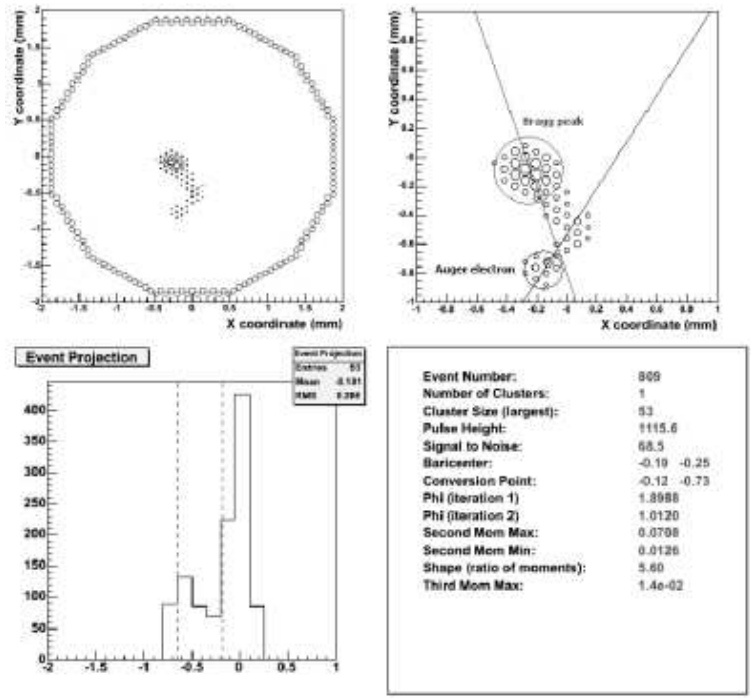}
\caption{Reconstructed track of a 5 KeV photoelectron. Track direction reconstruction algorithm: red line, first step;
blue line, second step.  (Read-out frequency 5MHz)}
\label{fig:12}
\end{center}
\end{figure}

\begin{figure}[ht]
\begin{center}
\includegraphics[width=10cm]{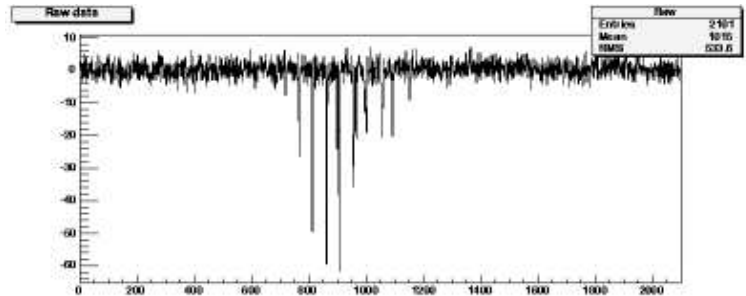}
\caption{The raw signals relative to the event shown in fig.\ref{fig:12}}
\label{fig:13}
\end{center}
\end{figure}

\begin{figure}[ht]
\begin{center}
\includegraphics[width=10cm]{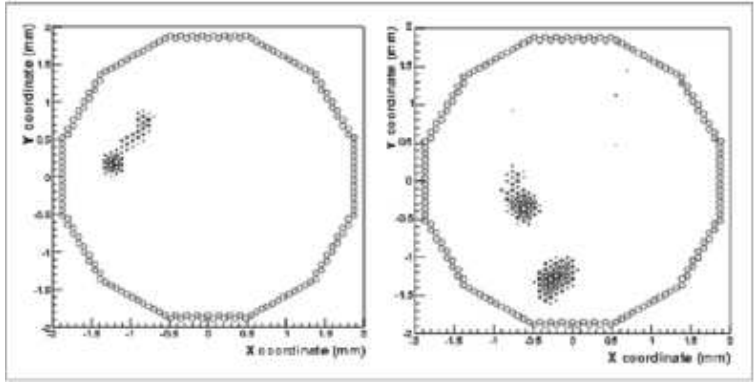}
\caption{Real reconstructed tracks}
\label{fig:14}
\end{center}
\end{figure}

A single GEM MPGD with an active gas volume of less than 1 cm$^{3}$ has been assembled directly over the chip die, so the ASIC itself has 
become the pixelized collecting anode of the detector. With this approach, for the first time, gas detectors have reached the level of 
integration and resolution typical of solid state pixel detectors. Different phases of the assembly are shown in fig.\ref{fig:7}.
In the actual prototype a drift region ({\em absorption gap}) of 6 mm above the GEM foil has been chosen, while a 1 mm spacer 
defines the {\em collection gap} between the bottom GEM and the pixel matrix of the read-out chip. The GEM has a standard thickness of
50 $\mu$m and holes of 50 $\mu$m at 90 $\mu$m pitch on a triangular pattern.  
The entrance window is a 25 $\mu$m Mylar foil, aluminized on one side. 
An artistic exploded view of the micro-mechanics of the detector is shown in fig.\ref{fig:8}. 
The gas mixture used to fill the detector is 80\% Neon, 20\% DME. Such a low Z gas mixture has been chosen for the high 
{\em stopping power/scattering ratio} and a still reasonable detection efficiency at low X-ray energy. Typical voltages applied to the drift 
electrode and to the GEM are respectively: -1000 V, -500 V (Top GEM), -100 V (Bottom GEM), the collecting electrodes being at $\sim$zero voltage.
In this condition the detector operates at a typical gain of 1000.
Thanks to the very low pixel capacitance at the preamplifier input, a noise level of 1.8 mV corresponding to $\sim$ 100 
electrons has been measured. The rms value of the pedestals distribution for each read-out channel is reported in fig.\ref{fig:9}. 
With a gas gain of 1000 and the measured noise level the detector has significant sensitivity to a single primary electron.
Strobing each pixel with 1 V signal ($\sim$1000 ADC counts) a uniformity of response of 3\% rms for all the 2101 channels has been observed 
(fig.\ref{fig:10}a). Because all the processing occurs within the pixel a negligible crosstalk has been measured in the channels 
adjacent to the ones pulsed with 1 V signal (see fig.\ref{fig:10}b).
The addressing capability of each individual pixel has been checked with the internal calibration system. The detector response to 20 mV
($\sim1000$ electrons) signal injected in a subset of pixels suitably chosen to create the experiment Logo is shown in fig.\ref{fig:11}. 
The first application of this new MPGD concept is for an Astronomical X-ray polarimeter in the low energy band 1$\div$10 keV. 
Information on the degree and angle of polarization of astronomical sources can be derived from the angular 
distribution of the initial part of the photoelectron tracks when projected onto a finely segmented 2D imaging detector.
As reported in previous papers (\cite{Quote3},\cite {Quote4}) the algorithm for the reconstruction of the photoelectron path starts
from the evaluation of the barycenter of the charge distribution on the read-out pixels and the maximization of the second moment 
({\em $M_{2}$}) of the charge distribution to define the principal axis of the track. In a further step the asymmetry of the charge release  
along the principal axis (third moment {\em $M_{3}$}) is computed and the conversion point derived by moving along this axis in the 
direction of negative {\em $M_{3}$}, where the released charge is smaller, by a length $\approx$ {\em $M_{2}$}. 
The reconstruction of the direction of emission is then done by
taking into account only the pixels in a region weighted according to the distance from the estimated conversion point.
The morphology of a {\em real} track obtained by illuminating the device with a low energy radioactive source 
(5.9 KeV X-ray from $^{55}$Fe) is shown in fig.\ref{fig:12}. 
The small cluster due to the Auger electron and the initial part of the track are well distinguishable from the larger Bragg 
peak. The projection of the charge distribution along the principal axis is also shown. The plot of the raw signals of all the channels for 
the same event shows the optimal signal to noise ratio obtained with this detector (fig.\ref{fig:13}). Around 50000 electrons from the 
gas amplified primary photoelectrons are subdivided on 53 pixels. 
Two {\em real} events, included a double track, are shown in Fig.\ref{fig:14}. 

\section{Conclusions}

A system in which the GEM foil, the absorption gap and the entrance window are assembled directly over a custom CMOS chip die has been 
developed. The transfer of  charge from the amplifying region to the collection and read-out region occurs via electric fields.
The ASIC itself becomes at the same time, the charge collecting anode and the pixelized read-out of a MicroPattern Gas Detector. 
For the first time the full electronics chain and the detector are completely integrated without the need of complicated bump-bonding.
At a gain of 1000 a high sensitivity to single primary electron detection is reached. No problems have been found up to now in 
operating the system under HV and in a gas environment. 
An astronomical X-ray Polarimeter application has been presented. Final design will have 16$\div$32 k channels and 60$\div$70 microns pixel size ($\simeq 1 cm^{2}$ active area). 
Depending on pixel and die size, electronics shaping time, analog vs. 
digital read-out, counting vs. integrating mode, gas filling, many others applications can be envisaged. 
This would open new directions in gas detector read-out, bringing the field to the same level of integration of solid state detectors.

\end{document}